\documentclass[conference]{IEEEtran}
\IEEEoverridecommandlockouts
\usepackage{cite}
\usepackage{amsmath,amssymb,amsfonts}
\usepackage{algorithmic}
\usepackage{graphicx}
\usepackage{textcomp}
\usepackage{xcolor}
\usepackage{bm}
\usepackage{multirow}
\usepackage{tabularx, colortbl, makecell}
\usepackage{etoolbox}
\usepackage{hyperref}

\makeatletter
\patchcmd{\@makecaption}
  {\scshape}
  {}
  {}
  {}
\makeatother

\def\BibTeX{{\rm B\kern-.05em{\sc i\kern-.025em b}\kern-.08em
    T\kern-.1667em\lower.7ex\hbox{E}\kern-.125emX}}
\begin{document}

\title{TSELM: Target Speaker Extraction using Discrete Tokens and Language Models
}

\author{\IEEEauthorblockN{1\textsuperscript{st} Beilong Tang}
\IEEEauthorblockA{\textit{Duke Kunshan University} \\
Kunshan, China \\
bt132@duke.edu}
\and
\IEEEauthorblockN{2\textsuperscript{nd} Bang Zeng}
\IEEEauthorblockA{\textit{Duke Kunshan University} \\
Kunshan, China \\
zeng.bang@dukekunshan.edu.cn}
% \and
% \IEEEauthorblockN{3\textsuperscript{rd} Zhan Jin}
% \IEEEauthorblockA{\textit{Duke Kunshan University} \\
% Kunshan, China \\
% zhan.jin@whu.edu.cn}
\and
\IEEEauthorblockN{3\textsuperscript{rd} Ming Li}
\IEEEauthorblockA{\textit{Duke Kunshan University} \\
Kunshan, China \\
ming.li369@dukekunshan.edu.cn}
}

\maketitle

\begin{abstract}
  We propose TSELM, a novel target speaker extraction network that leverages discrete tokens and language models. TSELM utilizes multiple discretized layers from WavLM as input tokens and incorporates cross-attention mechanisms to integrate target speaker information. Language models are employed to capture the sequence dependencies, while a scalable HiFi-GAN is used to reconstruct the audio from the tokens. By applying a cross-entropy loss, TSELM models the probability distribution of output tokens, thus converting the complex regression problem of audio generation into a classification task.
  Experimental results show that TSELM achieves excellent results in speech 
  quality and comparable results in speech intelligibility. 
\end{abstract}

\begin{figure*}[t]
    \centering
    \includegraphics[width=0.77\textwidth]{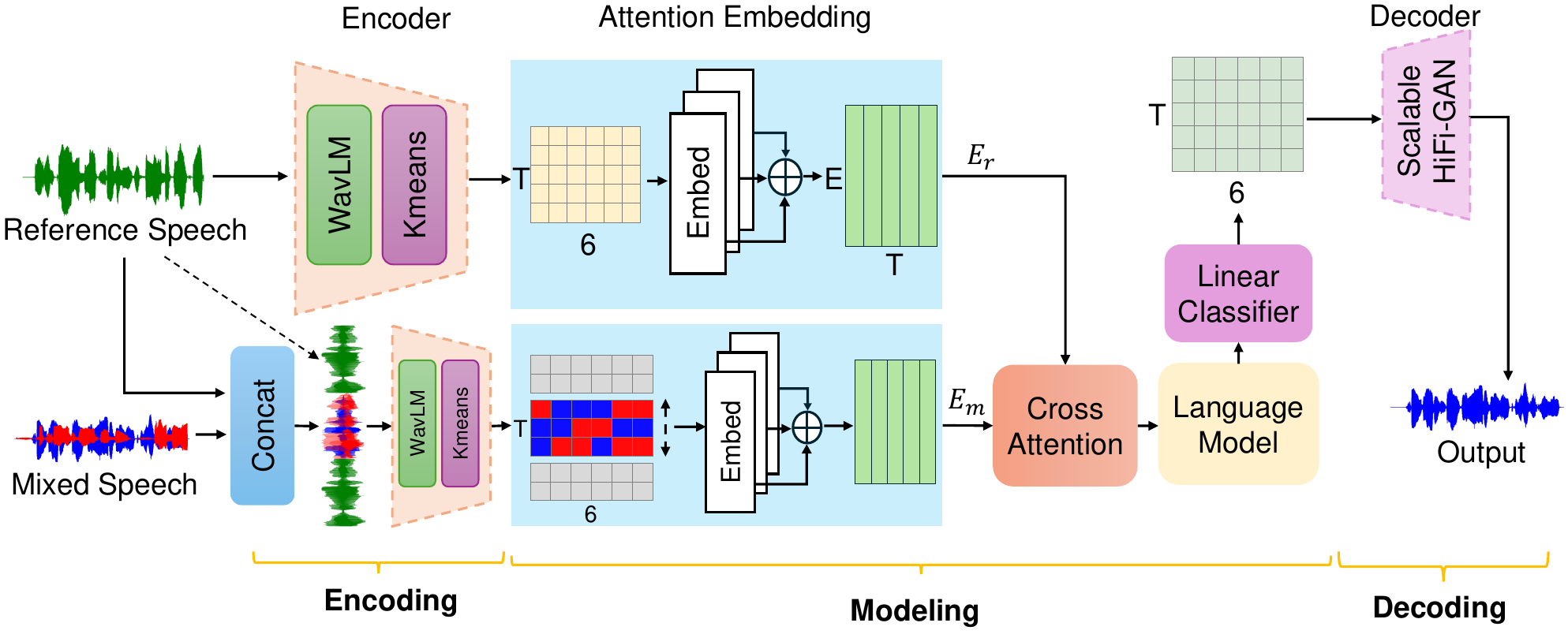}
    \caption{Overview of our proposed target speaker extraction framework with discrete tokens and language models.}
    \label{model}
    \vspace{-4pt}
    \end{figure*}

    \begin{figure}
        \centering
        \includegraphics[width=0.25\textwidth]{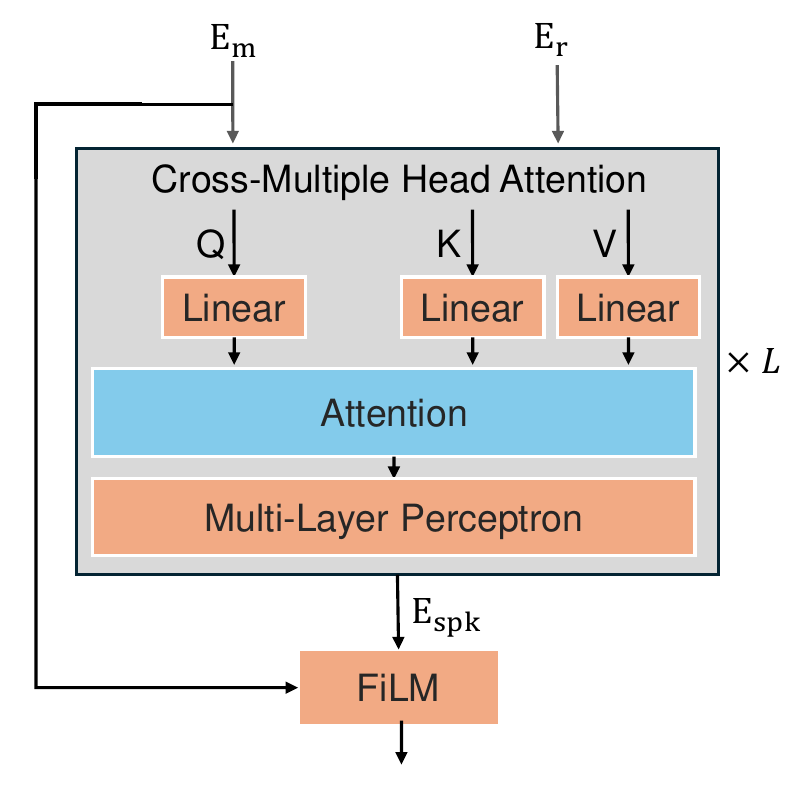}
        \caption{Details of the Cross-Attention mechanism in modeling.}
        \label{cross_attention}
        \vspace{-5pt}
        \end{figure}

\section{Introduction}

In contrast to blind speech separation, which seeks to isolate individual utterances from a mixture of known speakers, Target Speaker Extraction (TSE) focuses on extracting only the voice of the target speaker using auxiliary information. Current models are predominantly discriminative, employing masking strategies to directly minimize the distance between the estimated and clean speech signals \cite{luo2019conv,spex_plus,sepformer,sef_net}. However, these discriminative approaches often struggle to generalize to unseen data and may introduce undesirable distortions \cite{distortion}. To solve these issues, researchers have proposed generative models. This method aims to learn the underlying distribution of the target speaker's voice and use this knowledge to generate the clean speech of the target speaker from a mixture of voices rather 
than directly mapping from mixed speech to clean speech. Some generative models, like diffusion
models \cite{target_diff} and variational autoencoders (VAE) \cite{vae} have been studied. It has been demonstrated that generative models can achieve results comparable to those of discriminative models \cite{target_diff,tokensplit}.

The modeling of discretization of audio has gained significant attention with the advancement of language models (LMs) in modeling texts \cite{l1, l2, l3, l4, l5}. Researchers are trying to convert audio into discrete tokens and leverage LMs to 
model them, thereby simplifying audio generation tasks by transforming complex regression problems into classification tasks \cite{dasb}.
LMs have demonstrated scalability and unification of different modalities in audio synthesis tasks \cite{audio_gen,deshmukh2023pengi,chen2023lauragpt}. It has also achieved superior performances compared with discriminative models in
tasks such as speech enhancement \cite{selm,mask_sr} and blind speech separation \cite{tokensplit}.

Some discretization methods \cite{selm,tokensplit,dasb} use Kmeans on the output of Self-Supervised Learning (SSL) models such as HuBERT \cite{hubert} and WavLM \cite{wavlm}. SSL models have demonstrated outstanding performances across numerous downstream tasks \cite{superb}, as they extract continuous representations rich in semantic and timbral information from speech.

There are only few works about discretization in Target Speech Extraction (TSE). SkiM-UniCATS \cite{gen_tse} is among the first to use discrete tokens in TSE, leveraging SSL models like HuBERT and vq-wav2vec \cite{vq_wav2vec}. However, it overlooks the WavLM model, which excels in speech separation \cite{wavlm}. Also, it mainly focuses on single-layer outputs for discretization. Moreover, its evaluation is limited to speech quality, neglecting intelligibility and speaker similarity, both crucial for TSE. We try to address these potential issues by proposing TSELM, which uses multiple WavLM layers for discretization and adds metrics for intelligibility and speaker similarity. Our demos are available at 
\footnote{\href{https://beilong-tang.github.io/TSELM.demo/}{https://beilong-tang.github.io/TSELM.demo/}}.

\section{Method}
TSELM has three stages: encoding, modeling and decoding. In the encoding stage, both reference and mixed speech are tokenized using WavLM and Kmeans. The reference speech is passed directly to the encoder, while for the mixed speech, we concatenate the reference speech to both sides of the mixture before passing it through the WavLM model. After tokenization, we retain only the tokens corresponding to the mixed speech.
In the modeling stage, an attention embedding mechanism is employed to combine the embeddings from all layers. A cross-attention mechanism, similar to that used in \cite{usef_tes}, is applied to inject speaker-specific information. An encoder-only language model, followed by a linear classifier, is then used to generate the reconstructed tokens.
In the decoding stage, we leverage the pretrained scalable HiFi-GAN in \cite{unit_hifi} to reconstruct the audio from the discrete tokens. Unlike SELM \cite{selm}, where a conformer detokenizer is trained to reconstruct WavLM embeddings before passing them through HiFi-GAN, the scalable HiFi-GAN in \cite{unit_hifi} uses a dropout mechanism to directly reconstruct audio from multiple layers of tokens, eliminating the need for a conformer detokenizer and the complexity of training a separate HiFi-GAN for each layer. Both the encoder and decoder are kept frozen during training. An overview of TSELM is shown in Fig.\ref{model}.
Extensive experiments demonstrate that our method achieves excellent speech quality and comparable intelligibility results.

\subsection{Encoding}

We use the pretrained SSL model WavLM Large \cite{wavlm} to encode speech into continuous representations. Specifically, we extract the outputs from six hidden layers: 1, 3, 7, 12, 18, and 23. Given a speech signal \(s \in \mathbb{R}^{T'}\), the output of WavLM is a tensor \(\bm{r}\) with shape \(n \times T \times E\), where \(n\) is the number of output layers (6 in this case), \(T\) is the time dimension, and \(E\) represents the embedding dimension. For tokenization, we apply a separate Kmeans model to each output layer, with each model using the same number of clusters, denoted by \(K\). After tokenization, the continuous embedding \(\bm{r}\) is transformed into a discrete tensor \(\bm{d}\) with shape \(n \times T\), where each value \(\bm{d}_{i} \in (0, K-1) \). In all our experiments, we set \(K = 1000\). For both reference and mixed speech, the same Kmeans model and the same layer combination from WavLM Large are used. The encoder remains frozen during training.

The encoding strategy for mixed speech is crucial to the performance of the model. Given a reference speech \(s_r \in \mathbb{R}^{T^r}\) and a mixed speech \(s_m \in \mathbb{R}^{T'}\), we follow the previously described procedure to encode the reference speech into a tensor \(\bm{d_r}\) of shape \(n \times T_r\). However, for the mixed speech, instead of applying the encoding directly, we first concatenate it with the reference speech, creating a signal \(s' = [s_r, s_m, s_r] \in \mathbb{R}^{(T^r + T' + T^r)}\). This concatenated signal is then input into the encoder, producing an output tensor \(\bm{d'}\) with shape \(n \times (T_r + T + T_r)\), where \(T\) represents the length for the mixed speech embedding. The tensor \(\bm{d'}\) contains discrete tokens for the two segments of reference speech and the mixed speech. We extract the portion \(\bm{d}\) corresponding to the mixed speech, resulting in an output tensor of shape \(n \times T\).

This approach is inspired by WavLM \cite{wavlm}, which trains the model by overlapping clean speech with an interfering signal covering less than 50\% of the speech, using the first utterance as the primary speaker. This allows WavLM to focus on producing target speaker dependent embeddings. Our experiments demonstrate that this concatenation strategy significantly enhances the model's performance by guiding it to prioritize the target speaker's information.
\subsection{Modeling}

\begin{table*}
  \caption{Results on Libri2Mix clean and WSJ0-2mix: In the "Category" column, "G" refers to generative models, while "D" refers to discriminative models. The "Type" column categorizes methods as "D" (discrete), "H" (hybrid), or "C" (continuous).
  For discrete methods, speaker similarity is compared against the discretized target speech (denoted as Target-Discrete) instead of the original target speech. It is denoted as "\_d".}
  \renewcommand{\arraystretch}{1.2}
  \begin{center}
    \setlength{\tabcolsep}{4.9pt}
  \begin{tabular}{cccccccccccccccccc}
    \Xhline{2\arrayrulewidth} % Bold top line
  \multirow{3}{*}{System} & \multicolumn{1}{l}{\multirow{3}{*}{Category}} & \multicolumn{1}{l}{\multirow{3}{*}{Type}} &  \multicolumn{5}{c}{Libri2Mix Clean}                    &                               & \multicolumn{5}{c}{WSJ0\_2mix}                                                  \\
  \cline{4-8} \cline{10-14}
                          & \multicolumn{1}{l}{}                                                 & \multicolumn{1}{l}{}                            & \multicolumn{3}{c}{DNSMOS $\uparrow$} & dWER $\downarrow$ & Spk Sim $\uparrow$ &  & \multicolumn{3}{c}{DNSMOS $\uparrow$} & dWER $\downarrow$ & Spk Sim $\uparrow$  \\ \cline{4-6} \cline{10-12}
                          & \multicolumn{1}{l}{}                                                    & \multicolumn{1}{l}{}                            & SIG         & BAK        & OVL        &                   &        &             & SIG         & BAK        & OVL        &                   &                    \\ \hline
  Mixture                 & -                                             & -                                                                                           & 3.38        & 3.10       & 2.65       & 79.2            & -        &           & 3.42        & 3.28       & 2.81       & 63.6            & -                  \\
  Target-Discrete         & G                                             & -                                                                                          & 3.47       & 4.03       & 3.19       & 11.8            & 0.654     &          & 3.56        & 4.09       & 3.30       & 10.1            & 0.657               \\ \hline
  Spex+\cite{spex_plus}                   & D                                             & -                                                                                  & 3.38        & 3.77       & 3.00       & 19.0            & 0.922     &          & 3.49        & 4.00       & 3.21       & 15.0            & 0.943             \\
  SkiM-UniCATS(vq-wav2vec)\cite{gen_tse}                   & G                                             & D                                                                                  & -        & -      & -       & -            & -     &          & 3.62        & 4.10       & 3.37       & -            & -             \\ \hline
  Continuous-WavLM-L6     & G                                             & C                                                                                  & 3.57        & 4.06       & 3.28       & 14.6            & 0.870     &          & 3.61        & 4.08       & 3.35       & 8.0            & 0.892             \\
  TSELM-L-Hybrid          & G                                             & H                                                                                     & 3.49        & 4.05       & 3.22       & 20.0            & 0.917\_d  &          & 3.57        & 4.10       & 3.31       & 12.6            & 0.915\_d             \\
  TSELM-S-NoCat       & G                                             & D                                                                                   & 3.48        & 4.02       & 3.19       & 71.5            & 0.854\_d    &        & 3.55        & 4.08       & 3.28       & 64.5            & 0.888\_d         \\ \hline
  TSELM-S                 & G                                             & D                                                                                    & 3.50        & 4.06       & 3.23       & 28.1            & 0.883\_d      &      & 3.57        & 4.10       & 3.32       & 19.4            & 0.915\_d            \\
  TSELM-M                 & G                                             & D                                                                                   & 3.49        & 4.04       & 3.21       & 29.0            & 0.892\_d    &        & 3.57        & 4.10       & 3.32       & 18.8            & 0.921\_d             \\
  TSELM-L                 & G                                             & D                                                                                  & 3.49        & 4.04       & 3.21       & 27.5            & 0.895\_d    &        & 3.57        & 4.10       & 3.31       & 17.8            & 0.924\_d        \\
  \Xhline{2\arrayrulewidth} % Bold top line     
  \end{tabular}
  \label{main_exp}
\end{center}
\vspace{-8pt}
  \end{table*}

  \begin{table}
    \caption{Performance of different SSL models and layer selections on Libri2Mix clean. WavLM-L6 uses only the 6th layer of hidden output of the WavLM Large model. WavLM denotes our TSELM-L model which uses 6 hidden layers. 
            }
            \vspace{-15pt}
    \renewcommand{\arraystretch}{1.1}
    \begin{center}
      \setlength{\tabcolsep}{5pt}
        \begin{tabular}{ccccccc}
            \Xhline{2\arrayrulewidth} % Bold top line
            \multirow{2}{*}{SSL-Model} & \multirow{2}{*}{Type} & \multicolumn{3}{c}{ DNSMOS $\uparrow$} & \multirow{2}{*}{dWER $\downarrow$} & \multirow{2}{*}{Spk Sim $\uparrow$} \\
            \cline{3-5}
                                                       &                             & SIG     & BAK     & OVL    &                       &                          \\ 
            \hline
            \multirow{2}{*}{HuBERT}                              &  Discrete                          & 3.57    & 4.09    & 3.31   & 82.2                & 0.854\_d                        \\
                                             &  Hybrid                          & 3.57    & 4.10    & 3.32   & 36.1                & 0.900\_d                        \\
            \hline
            \multirow{2}{*}{WavLM-L6}                              &  Discrete                          & 2.08    & 2.07    & 1.64   & 122.14                & 0.589\_d                        \\
            &  Hybrid                          & 3.54    & 3.93    & 3.18   & 29.3                & 0.838\_d                        \\
            \hline
            \multirow{2}{*}{WavLM}                                     &  Discrete                          & 3.49    & 4.04    & 3.21   & 27.5                & 0.895\_d                        \\
                                              &  Hybrid                          & 3.49    & 4.05    & 3.22   & 20.0                & 0.917\_d                        \\
            
            \Xhline{2\arrayrulewidth} % Bold top line              
            \end{tabular}
            \linebreak
            \label{hubert_wavlm_6}
      \end{center}
      \vspace{-8pt}
    \end{table}

\subsubsection{Attention Embedding}
After obtaining the discrete tensor \(\bm{d}\) with shape \(6 \times T\), we use 6
learnable embedding tables each with \(K\) entires to embed the 6 layers 
respectively, each resulting in a tensor of shape \(T \times E\). 
After embedding, we follow the same recipe as in \cite{dasb} to 
aggregate the tensor by using attention mechanism to sum all the 6 tensors. This summation 
keeps the information of each layer while reducing the 
dimension of layers and system complexity. After attention embedding, we obtain reference embedding \(E_r\) and 
mixture embedding \(E_m\).

\subsubsection{Cross Attention}
We apply cross-attention module in \cite{usef_tes} to inject the reference embeddings into the mixture. The details 
are shown in Fig.\ref{cross_attention}.
The cross-attention module consists of a stack of cross-multiple head attention modules, followed by
a Feature-wise Linear Modulation (FiLM) module. 
We use \(E_m\) as the query and \(E_r\) as the key and value for 
the attention module. The output from the cross-multiple head attention module \(E_{spk}\)
is passed together with \(E_m\) to the FiLM to obtain the final output. The output of 
FiLM  \(E_f = FiLM(E_m, E_{spk}) = \gamma E_{spk} \cdot E_m  + \beta E_{spk} \) where 
\(\gamma\) and \(\beta\) are learnable parameters denoting the scaling and shifting vectors 
respectively.

\subsubsection{Language Modeling}
We use encoder-only transformers containing multiple self-attention modules to model the 
embedding \(E_f\). Due to encoder-only style, the LM is able to learn from all the positions. 
Finally, 6 linear classifiers each with dimension \(K\) is used to produce the logit scores of the tokens.
Cross-entropy loss is applied between the predicted tokens and the clean tokens, which are obtained by discretizing the ground truth clean audio.

\section{Experiments Setup}

\subsection{Training}

We use the publicly available Kmeans tokenizer and scalable 
HiFi-GAN decoder in \cite{speechbrain}. The Kmeans tokenizer
is trained on \texttt{train\_clean\_[100,360,500]} of
LibriSpeech \cite{librispeech}, and the scalable HiFi-GAN is trained on \texttt{train\_clean\_100} of 
LibriTTS \cite{libritts}. The modeling stage is trained on 
\texttt{train\_clean\_[100,360]} of LibriSpeech. All training data are 
generated on the fly with relative SNR between 0 to 5 dB. The mixture audio and reference audio is clipped to 3 and 4 seconds, respectively.

We utilize the output from hidden layers 1, 3, 7, 12, 18, 23 from WavLM Large and Kmeans model with 
\(K=1000\).
We use 1024 as the embedding dimension.  The cross-attention module consists of 4 transformer encoders, each with 16 attention heads and an MLP with a hidden dimension of 1024. Layer normalization is applied after the cross-attention module.
The LM of small version TSELM-S uses 
embedding dimension \(d\) = 256, absolute sinusoidal positional embedding, conformer encoders as the backbone of LM. The conformer encoder consists of 6 layers with a kernel size of 31, each with 4 attention heads, and an MLP with a hidden dimension of 2048. The medium version TSELM-M uses \(d\) = 512 with 8 layers and 8 heads and the large version TSELM-L uses 
\(d\) = 768 with 12 layers and 16 heads. We use AdamW as 
our optimizer for all the experiments. The learning rate 
is \(5 \times 10^{-4}\) for TSELM-S and \(5 \times 10^{-5}\) for TSELM-M and TSELM-L. We train our models using 8 GPUs with 32 GB of RAM, each with batch size 16 for a total of 40,000 steps.

\subsection{Evaluation}

We evaluate our models using Libri2Mix\cite{librimix} and WSJ0-2mix \cite{wsj0}.
We follow the recipes\footnote{\href{https://github.com/xuchenglin28/speaker_extraction}{https://github.com/xuchenglin28/speaker\_extraction}} to form the reference speeches for 
WSJ0-2mix.
We use the clean testset of Libri2Mix\footnote{\href{https://github.com/JorisCos/LibriMix}{https://github.com/JorisCos/LibriMix}}, and the reference speeches are randomly selected.

It is shown that metrics like PESQ, SI-SNR, STOI do not accurately reflect the speech quality of 
vocoder outputs due to the fact that the vocoder output does not focus strictly on frame alignment
\cite{tokensplit,selm}. Instead, we use DNSMOS \cite{dnsmos} to measure the speech quality, and the differential 
Word Error 
Rate (dWER) \cite{dwer} using the base model of Whisper \cite{whisper} to measure the speech intelligibility. For speaker similarity, we use the \texttt{ResNet221\_LM}
from the public WeSpeaker \cite{wespeaker} to calculate the embedding cosine speaker similarity.

\subsection{Baseline models}
Baseline models are presented in Table \ref{main_exp}. 
TSELM is compared with Spex+ \cite{spex_plus}, a discriminative separation model trained on Libri2Mix \cite{librimix}. 
We also compare our model with SkiM-UniCATS(vq-wav2vec) from \cite{gen_tse}, which uses discrete tokens from vq-wav2vec to conduct TSE. 
Mixture refers to the unprocessed mixed speech. 
Target-Discrete refers to the discretized target speech using our encoder and decoder. 
It serves as the upper bond for our model performance.  Besides TSELM, we conduct 3 main experiments, named
 Continuous-WavLM-L6, TSELM-L-Hybrid and TSELM-S-NoCat using the same training data. 
For Continuous-WavLM-L6, we directly
pass the embeddings from the 6th hidden layer output of WavLM Large to the cross-attention without discretization. 
The concatenation strategy is still applied to the mixed speech. Mean Square Error loss is applied between the output embeddings and the clean 
embeddings. HiFi-GAN in \cite{knn_vc} is used for audio reconstruction. For 
TSELM-L-Hybrid, inspired by MaskSR \cite{mask_sr}, we discretize the reference 
speech while utilizing the continuous embeddings from the mixed speech. In TSELM-S-NoCat, we utilize the TSELM-S model architecture but abandon the mixture concatenation strategy in the encoding step. The code and pretrained models are available at \footnote{\href{https://github.com/Beilong-Tang/TSELM}{https://github.com/Beilong-Tang/TSELM}}.

\section{Results and Discussions}
Table \ref{main_exp} presents the performance of different systems evaluated on the Libri2Mix testset and WSJ0-2mix testset. 
DNSMOS is computed over the 
output since this metric is reference-free. dWER is calculated with respect to the clean speech. 
For continuous methods, speaker similarity is directly computed against the clean speech. 
However, since discretization inherently results in some loss of speaker information—Target-Discrete shows a speaker similarity score of 0.654 on Libri2Mix —we follow \cite{dasb} to assess the speaker similarity of the output from the discrete methods against the target audio produced by discretizing the clean speech. 
The observed speaker information loss is likely due to the tokenization process, which inherently reduces speaker fidelity. Future work should aim to enhance tokenization methods for SSL models to mitigate this loss. Since the primary goal of this research is to explore target speaker extraction using discretized information rather than to develop improved tokenization methods, we argue that comparing the speaker similarity output with discretized clean speech is a reasonable approach, as it represents an upper bound of performance.

We observe that TSELM-L outperforms Spex+ in terms of DNSMOS scores, indicating better speech quality, but performs slightly worse in dWER, suggesting lower speech intelligibility. One potential reason for this could be the discretization process applied to the mixed speech. Our Kmeans algorithm is trained on clean speech rather than mixed speech, which is advantageous for speech enhancement as it likely aids in denoising. However, when applied to mixed speech, this discretization might lead to the model focusing on the wrong speaker, as it is most likely to retain only the dominant speaker information, causing a reduction in intelligibility.
This hypothesis is supported by the results of TSELM-L-Hybrid, where continuous embeddings from the mixture speech are used without discretization, achieving dWER scores better than Spex+ on WSJ0-2mix. 
% Another contributing factor could be the limitations of our current language model (LM). The encoder-only LM achieves around 55\% accuracy for token prediction on Libri2Mix testset, and we believe using more advanced models, such as auto-regressive or masking-based LMs, could lead to improved performance in future iterations.

We observe a significant increase in dWER when the mixture audio is not concatenated with the reference in the encoding step, as seen in the TSELM-S-NoCat results in Table \ref{main_exp}. For WavLM to effectively perform target speaker separation, the input audio must follow specific conditions: the mixture should be less than 50\% of the total length, and the first utterance should be the target speaker. Under these conditions, WavLM outputs a slightly denoised embedding that emphasizes the target speaker. When the entire input is a mixture, however, we found that WavLM sometimes extracts the wrong speaker.
Our current concatenation strategy, inspired by SELM's \cite{selm} success in speech denoising, reframes the target speaker extraction task as a more challenging speech enhancement problem, utilizing WavLM's denoising capabilities. However, this approach remains suboptimal, and we believe our future work should prioritize developing a speaker-aware tokenization method. 

In Table \ref{hubert_wavlm_6}, we compare the performance of HuBERT and WavLM as SSL models and examine the effects of using either one or multiple layers for discretization. Our findings indicate that using HuBERT as the SSL model results in slightly better DNSMOS scores but much worse dWER compared to our WavLM baseline. The improved DNSMOS scores likely stem from the vocoder performance, yet they do not adequately reflect speech intelligibility, which is crucial for speech separation tasks.
The poorer dWER scores observed with this HuBERT model may be attributed to its training on clean speech, which might not equip it to capture the complexity and richness of mixed speech. Moreover, our results from WavLM-L6 suggest that when performing speech separation, discretizing across multiple layers provides better results than relying on a single layer.  This might be because that using multiple layer outputs can better tolerate errors compared to using just one layer output.

Finally, we observe a performance gap between discrete methods and continuous methods, as demonstrated by Continuous-WavLM-L6.
Continuous-WavLM-L6 has the best performance in terms of DNSMOS and dWER among all the experiments while using only the 6th layer output of WavLM Large. 
The gap in performance may be attributed to the information loss inherent in the discretization process. 
We hope our future research will bridge this gap.   

\section{Conclusion}
In this work, we introduce a novel way using discrete tokens and language models for target speaker extraction.
Our method leverages multiple hidden layers of WavLM and Kmeans tokenizers for encoding, employs cross-attention and a language model for separation, and utilizes a scalable HiFi-GAN for audio reconstruction.
Experiments have 
shown that our model can achieve excellent performance in terms of speech quality, and comparable performance in terms of speech intelligibility. However, we observe a gap between discrete methods and continuous methods especially in speech intelligibility and speaker similarity. Our future work will focus on shrinking 
this gap.

% \section*{Acknowledgment}

% We want to thank for Kunshan Super Computing SCNet for providing the computing resources. 

\bibliographystyle{IEEEtran}
\bibliography{paper}

\end{document}